# The thermostatistical aspect of Werner-type states and quantum entanglement

Sumiyoshi Abe[1,2,4], A. R. Usha Devi[3,4] and A. K. Rajagopal[4]

[1] Department of Physical Engineering, Mie University, Mie 514-8507, Japan

[2] Institut Supérieur des Matériaux et Mécaniques Avancés, 44 F. A. Bartholdi, 72000 Le Mans, France

[3] Department of Physics, Bangalore University, Bangalore-560 056, India

[4] Inspire Institute Inc., McLean, Virginia 22101, USA

**Abstract**

A general Werner-type state is studied from two viewpoints: (i) an application of dynamical interaction of the objective system with its environment, represented by a unital positive operator-valued measure (POVM), which ensures increase of entropy, makes the system evolve from an initial pure state to a mixed state of the Werner type, and (ii) the space of the objective system is constrained to have a given value of the fidelity. Then, the maximum entropy principle is shown to yield the Werner-type state as a canonical ensemble with the projector Hamiltonian. This novel observation is illustrated by examples of bipartite systems, the separability criteria on which are given in terms of the values of temperature. The present viewpoint may cast light on relevance of thermostatistics to the physics of quantum entanglement. In addition, the POVM scheme presented here offers a way of experimentally generating the Werner-type thermal states.

In quantum information theory and technology [1], there are at least two issues of great current interest. One is to control decoherence and decay of a pure state to a mixed state. The other is to understand quantum entanglement and separability criteria for density matrices of multi-partite systems [2].

In this article, we study a special class of mixed states termed the Werner-type states and present a novel viewpoint that amalgamates the twin concepts of mixedness and separability. We consider "thermalization" of the open objective system based on a dynamical approach and replace the effects of the hidden environment with quantum operations represented by positive operator-valued measures (POVMs) [1]. We then show that both this scheme and the maximum entropy principle [3] with the constraint on the fidelity [4] lead to the Werner-type states. In this way, we are able to address the problems of both separability of a density matrix of the objective system and decay of a pure state to a mixed state from the statistical-mechanical point of view. We explicitly describe these concepts by employing the Werner state of the two-qubit system [5] and the $(2j+1) \times (2j+1)$ rotationally invariant states [6] as well.

Let us start our discussion with considering an arbitrary *n*-partite system as the objective system whose Hilbert space is *d*-dimensional. There are two particular states of this system: one is the pure state, $\rho_0 = |\psi\rangle\langle\psi|$ with $|\psi\rangle$ being a certain normalized *entangled* state vector in the Hilbert space (e.g., the ground state of a many-body system), the von Neumann entropy (discussed later) of which is zero, and the other is the completely mixed state, $\rho_1 = I/d$ ($I$ being the identity matrix) having the maximum value of the entropy. Then, construct the mixture of these two extreme states, i.e., their convex combination

$$\tilde{\rho} = x\rho_0 + (1-x)\rho_1, \tag{1}$$

where $x \in [0, 1]$. $\tilde{\rho}$ is referred to here as the Werner-type state with an entangled pure state, $\rho_0$. One parameter families of density matrices having the form in Eq. (1) play an important role in understanding the physics of quantum entanglement in mixed states. There may be a specific value of $x$, say $x_S$, below which $\tilde{\rho}$ is separable, that is, $\tilde{\rho}$ can be written as follows:

$$\tilde{\rho}_{\text{sep}} = \sum_\lambda p_\lambda \, \rho_\lambda^{(1)} \otimes \rho_\lambda^{(2)} \otimes \cdots \otimes \rho_\lambda^{(n)}, \tag{2}$$

where $\rho_\lambda^{(i)}$ is the state of the $i$th partite, $p_\lambda \in [0, 1]$, and $\sum_\lambda p_\lambda = 1$. A state, which is not separable, is said to be entangled. In practice, it is a difficult problem to find $x_S$ [2,7,8].

A state like that in Eq. (1) can experimentally be realized [9,10], and therefore it has a definite physical importance. We first observe that the density matrix in Eq. (1) is of a thermal state. This is because it can be recast in the following canonical form:

$$\tilde{\rho} = \frac{1}{Z(\beta)} e^{-\beta H}, \tag{3}$$

where $\beta$ is the *inverse temperature* $T^{-1}$ with the unit Boltzmann constant, and $H$ and $Z(\beta)$ are the projector Hamiltonian and the partition function given by

$$H = -g|\psi\rangle\langle\psi|, \tag{4}$$

$$\begin{aligned} Z(\beta) &= \mathrm{Tr}\, e^{-\beta H} \\ &= d - 1 + e^{\beta g}, \end{aligned} \tag{5}$$

respectively, where $g$ is taken to be a positive coupling constant with the dimension of the energy. A general consideration of exponential distributions as in Eq. (3) has been developed in Ref. [11].

Eq. (3) is, in fact, identical to Eq. (1) if the following identification is made:

$$x = \frac{e^{\beta g} - 1}{d - 1 + e^{\beta g}}. \tag{6}$$

Therefore, the value of the temperature corresponding to the separability point is

$$T_S = \frac{g}{\ln\left(1 + \dfrac{x_S d}{1 - x_S}\right)}, \tag{7}$$

provided that $x_S$ depends on $d$, in general. We mention that discussions about separabilities of thermal states were made in recent works in Refs. [12,13].

The equivalence between Eqs. (1) and (3) is quite straightforward. To our knowledge, however, this has not been noticed earlier in the literature.

We would like to make the following comments here. It is well known that the thermal state at zero temperature is determined by the lowest eigenstate of the system Hamiltonian, $H$, such that $H|\psi_0\rangle = E_0|\psi_0\rangle$ with $E_0$ being the ground-state energy. One may write the Hamiltonian in terms of the complete set of its complete orthonormal eigenstates as follows: $H = E_0|\psi_0\rangle\langle\psi_0| + \sum_{i\neq 0} E_i|\psi_i\rangle\langle\psi_i|$, where we have isolated the ground-state term for emphasis. In the case when the energies of the excited states are much larger than that of the ground state (such as particles confined in a very small container), the system is effectively described by the dynamics of the lowest mode, that is, the projector Hamiltonian $H \cong E_0|\psi_0\rangle\langle\psi_0|$. In this way, the pure state density matrix, $\rho_0 = |\psi\rangle\langle\psi|$ introduced above and the projector Hamiltonian in Eq. (4) (with $E_0 \equiv -g$) can physically be interpreted. Embedding such a system in the heat bath entails interaction, which converts the pure state into a mixed state. This state change may be represented by a non-unitary quantum operation, which is a "heat-up operation".

On the other hand, we observe that the thermalization of the pure state $\rho_0$ naturally arises from the maximum entropy principle [3] under the constraint on a value of the fidelity [4]. Such a constraint maintains "propensity" of the mixed state to be close to $|\psi\rangle$ as expressed in Eq. (9) below. More explicitly, let us consider the maximization of

the von Neumann entropy

$$S[\rho] = -\text{Tr}(\rho \ln \rho) \tag{8}$$

under the constraints on the expectation value of the projector Hamiltonian, which is in fact the fidelity (see Refs. [14-16])

$$F = \langle \psi | \rho | \psi \rangle \tag{9}$$

as well as the normalization condition, $\text{Tr}\rho = 1$. The variation of the functional

$$\Sigma[\rho: \alpha, \beta] = S[\rho] - \alpha(\text{Tr}\,\rho - 1) + \beta g\left(\langle \psi | \rho | \psi \rangle - F\right) \tag{10}$$

with respect to $\rho$ yields the state in Eq. (1), where $\alpha$ and $\beta$ are Lagrange's multipliers, and $\alpha$ is to be eliminated by the normalization condition. The result is

$$\tilde{\rho} = \frac{Fd-1}{d-1} |\psi\rangle\langle\psi| + \frac{1-F}{d-1} I, \tag{11}$$

where

$$F = \frac{e^{\beta g}}{d-1+e^{\beta g}}$$

$$= \frac{1+(d-1)x}{d}. \tag{12}$$

Now, we consider the effect of the hidden environment on the objective system as a linear, positive, and trace-preserving map. Such a map is given in terms of the set of operators, $\{V_k\}$, as follows:

$$\rho \to \Phi(\rho) = \sum_k V_k \rho V_k^\dagger, \qquad (13)$$

with the trace-preserving condition

$$\sum_k V_k^\dagger V_k = I. \qquad (14)$$

The set $\{V_k^\dagger V_k\}_k$ is a positive operator-valued measure (POVM). What we are interested in here is a subclass of POVM termed unital POVM [17,18], which satisfies not only Eq. (14) but also the following condition:

$$\sum_k V_k V_k^\dagger = I. \qquad (15)$$

This map has two physically important properties, viz. (a) the completely mixed state is its fixed point:

$$\Phi(I) = I, \qquad (16)$$

and (b) it ensures increase of the entropy [17-19]:

$$S[\Phi(\rho)] \geq S[\rho]. \tag{17}$$

Thus, a unital POVM maps a pure state, $\rho_0 = |\psi\rangle\langle\psi|$, to the completely mixed state, $\rho_1 = I/d$, with the monotonic increase of the entropy. It should be noticed that there is freedom in representing such an operation.

However, the question is if such an operation varies the pure state to the completely mixed state *along* the curve of the convex combination in Eq. (1). It seems that there is no general answer to this question. However, below, we would like to present simple but nontrivial examples that show that it is indeed the case.

The first example is the Werner state [5] of a two-qubit system, $(A, B)$. The Heisenberg Hamiltonian describing such a system reads $H = J\sigma_A \cdot \sigma_B$ with a positive coupling constant $J$, the ground state of which is $\rho_0 = |\Psi^-\rangle\langle\Psi^-|$ with the Bell state $|\Psi^-\rangle = (|1\rangle^{(A)}|0\rangle^{(B)} - |0\rangle^{(A)}|1\rangle^{(B)})/\sqrt{2}$ having the energy $-3J$. The corresponding thermal state is given by

$$\tilde{\rho}_W(x) = x|\Psi^-\rangle\langle\Psi^-| + \frac{1-x}{4} I^{(A)} \otimes I^{(B)}, \tag{18}$$

which is in fact the Werner state, where $I^{(A)}$ ($I^{(B)}$) is the identity matrix of $A$ ($B$) and $x \in [0, 1]$. The fidelity of this state with respect to $|\Psi^-\rangle$ is given by $F_W = (1 + 3x)/4$, and the separability point of this state is at $x_S = 1/3$ [7]. Consider the following unital POVM:

$$V_0(x) = \frac{\sqrt{1+3x}}{2} I^{(A)} \otimes I^{(B)}, \qquad V_1(x) = \frac{\sqrt{1-x}}{2} I^{(A)} \otimes \sigma_x^{(B)},$$

$$V_2(x) = \frac{\sqrt{1-x}}{2} I^{(A)} \otimes \sigma_y^{(B)}, \qquad V_3(x) = \frac{\sqrt{1-x}}{2} I^{(A)} \otimes \sigma_z^{(B)}, \tag{19}$$

or the interchange, $A \leftrightarrow B$, of these, where $\sigma$'s are the Pauli matrices. The map, $\Phi_x$, associated with this POVM takes the "initial" pure Bell state, $\rho_0 = |\Psi^-\rangle\langle\Psi^-|$, to the mixed state, $\tilde{\rho}_W$:

$$\rho_0 \to \Phi_x(\rho_0) = \sum_{k=0}^{3} V_k(x) \rho_0 V_k^\dagger(x) = \tilde{\rho}_W(x). \tag{20}$$

This can be seen as a heat-up operation, and $\{\Phi_x \mid x \in [0,1]\}$ forms a one-parameter Abelian semigroup. A successive application of this map yields

$$\left(\cdots * \Phi_{x_3} * \Phi_{x_2} * \Phi_{x_1}\right)(\rho_0) = \tilde{\rho}_W(x_1 x_2 x_3 \cdots). \tag{21}$$

Clearly, such a successive operation can traverse through the separation point, $x_S = 1/3$, corresponding to the temperature, $T_S = g/\ln 3$. Ultimately, it transforms the pure Bell state, $\rho_0 = |\Psi^-\rangle\langle\Psi^-|$, to the completely mixed state, $\rho_1 = I_A \otimes I_B / 4$, unless $x_i = 1$ ($i = 1, 2, 3, ...$). Thus, we have established the thermalization of the Bell state.

The above discussion has an immediate implication towards experimentally generating Werner-type states, with different fidelities, via the POVM scheme illustrated, which is different from those given in Refs. [9,10].

We present an interpretation of the above results by considering the coupled unitary evolution of the objective system and the environment (which is initially uncorrelated) resulting in a general scheme of quantum dynamical operation: $\rho_{\text{system}} \to \Phi(\rho_{\text{system}}) = \text{Tr}_{\text{env}}[U\rho_{\text{system}} \otimes \rho_{\text{env}} U^{\dagger}] = \sum_k V_k \rho_{\text{system}} V_k^{\dagger}$. Take an initially uncorrelated pure state of the $N$ bipartite systems, $|\chi\rangle = |\Psi^-\rangle_1 |\Psi^-\rangle_2 \cdots |\Psi^-\rangle_N$. Here, one of the Bell states, say, $|\Psi^-\rangle_1$, is regarded as the state of the objective system and the rest play the role of the environment. A nonlocal unitary transformation, $U$, associated with a POVM map on the state of the objective system, $|\Psi^-\rangle_1$, leads to an entangled state of the total system:

$$U(x)|\chi\rangle = \sum_{k=0}^{3} c_k(x) |\phi^{(k)}\rangle_1 |\phi^{(k)}\rangle_2 \cdots |\phi^{(k)}\rangle_N, \qquad (22)$$

where $\{|\phi^{(k)}\rangle\}_{k=0,1,2,3} \equiv \{|\Psi^-\rangle, |\Psi^+\rangle, |\Phi^-\rangle, |\Phi^+\rangle\}$ are the Bell basis, $|\Psi^\pm\rangle = (|1\rangle|0\rangle \pm |0\rangle|1\rangle)/\sqrt{2}$, $|\Phi^\pm\rangle = (|1\rangle|1\rangle \pm |0\rangle|0\rangle)/\sqrt{2}$, and

$$c_0(x) = \frac{\sqrt{1+3x}}{2}, \quad c_1(x) = c_2(x) = c_3(x) = \frac{\sqrt{1-x}}{2} \qquad (23)$$

are the Schmidt coefficients, which characterize system-environment entanglement. Upon tracing over the environmental degree of freedom, the state of the objective system experiences a quantum dynamical operation and becomes thermal as desired:

$$\tilde{\rho}_{1W}(x) = \text{Tr}_{\text{env}}[U(x)|\chi\rangle\langle\chi|U^{\dagger}(x)] = \sum_{k=0}^{3} c_k^2(x) |\phi^{(k)}\rangle^{(1)\,(1)}\langle\phi^{(k)}|. \qquad (24)$$

In a particular case when $c_0 = c_1 = c_2 = c_3 = 1/2$, the system is in the completely mixed state. This reveals that the iterative unital POVM map has its fixed point as the completely mixed state, corresponding to the maximally entangled state of the total system.

Another example we discuss here involves the rotationally invariant state [6] of a two-qudit system (A, B) with $d = 2j+1$. The state is defined by

$$\tilde{\rho}_{\text{rot-inv}}(x) = x \, |(jj)\,00\rangle\langle(jj)\,00| + \frac{1-x}{(2j+1)^2} I^{(A)} \otimes I^{(B)} \qquad (25)$$

with $x \in (0, 1)$, where the bipartite singlet state $|(jj)\,00\rangle$ is given in terms of the subsystem angular momentum states, $|j, m\rangle$, as

$$|(jj)\,00\rangle \equiv \frac{1}{\sqrt{2j+1}} \sum_{m=-j}^{j} (-)^{j-m} |j, m\rangle^{(A)} |j, -m\rangle^{(B)}. \qquad (26)$$

The fidelity of the state in Eq. (25) with respect to $|(jj)\,00\rangle$ is given by $F_{\text{rot-inv}} = (1-x)/(2j+1)^2 + x$. Its concurrence is related to the fidelity by [20]

$$C(\tilde{\rho}_{\text{rot-inv}}) = \begin{cases} 0 & (F < 1/(2j+1)) \\ \sqrt{(2j+1)/j}\,[F - 1/(2j+1)] & (1/(2j+1) \leq F \leq 1) \end{cases}, \qquad (27)$$

and the state is separable if $F \leq 1/(2j+1)$ [21]. Clearly, the state in Eq. (25) is a thermal state, $\tilde{\rho}_{\text{rot-inv}} = e^{-\beta H}/Z(\beta)$, with $H = -g|(jj)\,00\rangle\langle(jj)\,00|$ being the lowest mode of the Heisenberg Hamiltonian of a $j$-state system, where $Z(\beta) = (2j+1)^2 - 1 + e^{\beta g}$ and $F = e^{\beta g}/Z(\beta)$. The separation temperature reads $T_S = g/\ln(2j+2)$.

We now construct a unital POVM for the above-mentioned state as follows. Consider a complete orthonormal set, $\{|(jj)JM\rangle | J = 0, 1, ..., 2j; -J \leq M \leq M\}$, in the Hilbert space of the bipartite system. The operator defined by

$$v_{JM} = I^{(A)} \otimes I^{(B)} + |(jj)JM\rangle\langle(jj)00| + |(jj)00\rangle\langle(jj)JM|$$
$$- |(jj)00\rangle\langle(jj)00| - |(jj)JM\rangle\langle(jj)JM| \qquad (28)$$

possesses the following properties: $v_{JM}^\dagger = v_{JM}$, $v_{00} = I^{(A)} \otimes I^{(B)}$, $v_{JM}^\dagger v_{JM} = v_{JM} v_{JM}^\dagger = I^{(A)} \otimes I^{(B)}$, and $v_{JM}|(jj)00\rangle = |(jj)JM\rangle$. The required unital POVM is then given in terms of

$$\{V_{JM} = a_{JM} v_{JM} | J = 0, 1, ..., 2j; -J \leq M \leq J\}, \qquad (29)$$

where $a_{JM}$'s are real coefficients satisfying $\sum_{J,M} a_{JM}^2 = 1$. With a further choice of the parameters

$$a_{00} = a(x) = \frac{\sqrt{1 + [(2j+1)^2 - 1]x}}{2j+1}, \quad a_{JM} = b(x) = \frac{\sqrt{1-x}}{2j+1}$$
$$(J = 1, 2, ..., 2j; -J \leq M \leq J) \qquad (30)$$

such that $a^2(x) + [(2j+1)^2 - 1]b^2(x) = 1$, we find that the map, $\Phi_x$, associated with this unital POVM transforms the pure state, $\rho_0 = |(jj)\,00\rangle\langle (jj)\,00|$, to the thermal state in Eq. (25), as desired:

$$\rho_0 \to \Phi_x(\rho_0) = \sum_{J,M} V_{JM}(x)\,\rho_0\,V_{JM}^{\dagger}(x) = \tilde{\rho}_{\text{rot-inv}}(x). \qquad (31)$$

As in the preceding two-qubit example, the set $\{\Phi_x \mid x \in [0,1]\}$ forms an Abelian semigroup, and a successive application of the map $\Phi_x$ traverses through the separation point, finally reaching the fixed-point completely mixed state, in the same manner as illustrated in Eq. (21). Also, a nonlocal unitary transformation of the state of the total system-environment state, $|\chi\rangle = |(jj)\,00\rangle_1 |(jj)\,00\rangle_2 \cdots |(jj)\,00\rangle_N$ with $|(jj)\,00\rangle_1$ representing the state of the objective system and the rest forming the environment, such that $U(x)|\chi\rangle = \sum_{J=0}^{2j}\sum_{M=-J}^{J} a_{JM}(x) |(jj)\,JM\rangle^{(1)} |(jj)\,JM\rangle^{(2)} \cdots |(jj)\,JM\rangle^{(N)}$, gives the thermal state in Eq. (25) as the state of the objective system, when partial trace over the environmental degrees of freedom is performed, as in the previous example. The eigenvalues, $a^2(x)$ and $b^2(x)$ [with degeneracy $(2j+1)^2 - 1$] of $\tilde{\rho}_{\text{rot-inv}}(x)$ thus get interpreted as the Schmidt coefficients characterizing system-environment entanglement.

Finally, we would like to point out that the present scheme may be accomplished experimentally. In fact, possible experimental implementation of POVM operations on physical systems has recently been gaining attention [22].

To summarize, starting from the observation that a unital POVM map transforms a pure state to a mixed state, we have constructed the Werner-type states and shown in view of the maximum entropy principle with the projector Hamiltonians that they are actually thermal states. In particular, we have given details of the scheme for explicit examples of the two-qubit and two-qudit systems embedded in the symmetric environments and discussed their separability conditions in terms of the values of temperature. A confluence of three concepts, viz. the maximum entropy principle, unital dynamical POVM maps, and system-environment entanglement, play a unified role together in establishing thermalization of pure states to the Werner-type states. As already emphasized earlier, the present discussion may cast new light on possible roles of thermostatistics in deeper understanding of the physics of quantum entanglement [23].

**Acknowledgment**

S. A. was supported in part by a Grant-in-Aid for Scientific Research from the Japan Society for the Promotion of Science.

**References**

[1]  Nielsen M A and Chuang I L 2000 *Quantum Computation and Quantum Information* (Cambridge: Cambridge University Press)

[2]  Horodecki R, Horodecki P, Horodecki M and Horodecki K 2007 *Preprint*

quant-ph/0702225

[3] Rosenkrants R D (ed) 1989 *E. T. Jaynes: Papers on Probability, Statistics and Statistical Physics* (Dordrecht: Kluwer)

[4] Jozsa R 1994 *J. Mod. Opt.* **41** 2315

[5] Werner R F 1989 *Phys. Rev.* A **40** 4277

[6] Schliemann J 2003 *Phys. Rev.* A **68** 012309

[7] Peres A 1996 *Phys. Rev. Lett.* **77** 1413

[8] Horodecki M, Horodecki P and Horodecki R 1996 *Phys. Lett.* A **223** 1

[9] Zhang Y-S, Huang Y-F, Li C-F and Guo G-C 2002 *Phys. Rev.* A **66** 062315

[10] Agarwal G S and Kapale K T 2006 *Phys. Rev.* A **73** 022315

[11] Brody D C 2007 *J. Phys. A: Math. Theor.* **40** F691

[12] Khemmani S, Sa-Yakanit V and Strunz W T 2005 *Phys. Lett.* A **341** 87

[13] Markham D, Anders J, Vedral V, Murao M and Miyake A 2008 *Europhys. Lett.* **81** 40006

[14] Horodecki R, Horodecki M and Horodecki P 1999 *Phys. Rev.* A **59** 1799

[15] Abe S and Rajagopal A K 1999 *Phys. Rev.* A **60** 3461

[16] Rajagopal A K 1999 *Phys. Rev.* A **60** 4338

[17] Chefles A 2002 *Phys. Rev.* A **65** 052314

[18] Abe S and Rajagopal A K 2003 *Phys. Rev. Lett.* **91** 120601

[19] Bhatia R 1997 *Matrix Analysis* (New York: Springer-Verlag)

[20] Rungta P and Caves C M 2003 *Phys. Rev.* A **67** 012307

[21] Horodecki M and Horodecki P 1999 *Phys. Rev.* A **59** 4206

[22] Ahnert S E and Payne M C 2006 *Phys. Rev.* A **73** 022333

[23] Vedral V 2006 *Introduction to Quantum Information Science*

(Oxford: Oxford University Press)